\documentclass[%
 aip,
 amsmath,amssymb,
 reprint,%
]{revtex4-1}

\usepackage{graphicx}
\usepackage{dcolumn}
\usepackage{bm}

\usepackage{color,soul}
\setstcolor{red}
\usepackage{balance}

\newcommand{\RN}[1]{%
  \textup{\uppercase\expandafter{\romannumeral#1}}%
}

\usepackage[utf8]{inputenc}
\usepackage[T1]{fontenc}
\usepackage{mathptmx}
\usepackage{etoolbox}
\usepackage{multirow}
\usepackage{xcolor}

\makeatletter
\def\@email#1#2{%
 \endgroup
 \patchcmd{\titleblock@produce}
  {\frontmatter@RRAPformat}
  {\frontmatter@RRAPformat{\produce@RRAP{*#1\href{mailto:#2}{#2}}}\frontmatter@RRAPformat}
  {}{}
}%
\makeatother
\begin{document}

\preprint{AIP/123-QED}

\title{Constant $di/dz$ Scanning Tunneling Microscopy: Atomic Precision Imaging and Hydrogen Depassivation Lithography on a $\mathrm{Si(100)-2\times 1:H}$ Surface}

\author{Richa Mishra}
\author{S. O. Reza Moheimani}%
 \email[Corresponding author: ]{reza.moheimani@utdallas.edu}
\affiliation{Erik Jonsson School of Engineering and Computer Science, The University of Texas at Dallas, Richardson, Texas 75080, USA
}
\date{\today}

\begin{abstract}
We introduce a novel control mode for Scanning Tunneling Microscopy (STM) that leverages $di/dz$ feedback. By superimposing a high-frequency sinusoidal modulation on the control signal, we extract the amplitude of the resulting tunneling current to obtain a $di/dz$ measurement as the tip is scanned over the surface. A feedback control loop is then closed to maintain a constant $di/dz$, enhancing the sensitivity of the tip to subtle surface variations throughout a scan. This approach offers distinct advantages over conventional constant-current imaging. We demonstrate the effectiveness of this technique through high-resolution imaging and lithographic experiments on several Si(100)-2x1:H surfaces. Our findings, validated across multiple STM systems and imaging conditions, pave the way for a new paradigm in STM control, imaging, and lithography.

\end{abstract}

\maketitle

\section{\label{sec:intro}Introduction}
The Scanning Tunneling Microscope (STM) is a powerful instrument that provides 3-D real-space images and allows spatially localized measurements of both geometric and electronic structures of a conducting surface \cite{binning1982surface,stroscio1986electronic}. Due to its remarkable precision and control over individual atoms, it has been used in emerging atomically precise manufacturing and lithography applications \cite{ballard2013multimode, lyding1994nanoscale}. The STM holds immense promise for fabricating materials and devices with unparalleled accuracy and intricacy \cite{adams2016harnessing,avouris1996stm,lee2014lithography}.

\begin{figure}[h!]
\centering{}
\includegraphics[scale=0.5]{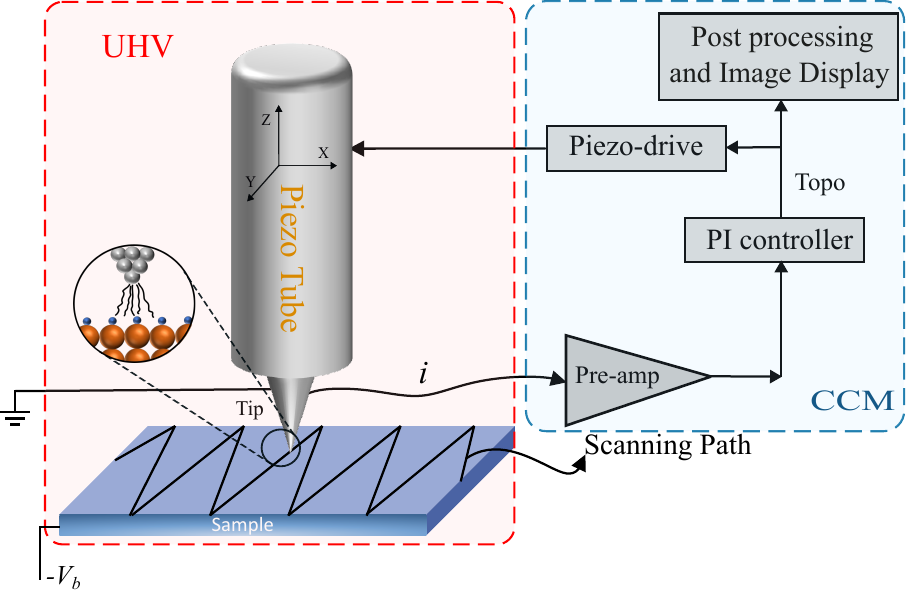}
\caption{Schematics of the scanning tunneling microscope working in conventional constant current mode.}
\label{f:sch_lnI}
\end{figure}

The functionality of this device is based on the quantum phenomenon known as tunneling current \cite{binnig2000scanning}. In this process, when two electrically conductive surfaces come into proximity by a distance less than 1\,$nm$, and are subjected to a voltage bias, electrons surpass the potential barrier and generate a highly sensitive tunneling current. By maintaining a constant intensity of this tunneling current during a scan, it is possible to capture images of surface variations with angstrom-level resolution \cite{binnig19837} as shown in Fig.\,\ref{f:sch_lnI}.

Piezoelectric actuators are integrated with the STM for precise movement in three dimensions and accurate scanning control. However, achieving sub-nanometer scale precision is challenging due to system nonlinearities and noise \cite{cao2015survey, blanvillain2013subnanometer}. Another critical parameter in STM scanner design is the resonant frequency of the piezoelectric tube scanner, which limits the scan rate, dictates stability against mechanical vibrations, and influences strategies for electronic noise reduction. The limitations of STM control systems often lead to poor topography images and resolution \cite{anguiano1998optimal}. Despite its importance, only a limited number of efforts have been made to enhance the STM control system since its invention over forty years ago.

An analysis of the STM control system was conducted by \citeauthor{oliva1995analysis}\cite{oliva1995analysis}, focusing on determining optimal imaging conditions and identifying the most suitable feedback parameters \cite{oliva1997experimental}. Further efforts were made to design robust controllers specifically tailored to a given STM, as detailed in the studies by \citeauthor{ahmad2012robust}\cite{ahmad2012robust}. To prevent STM tip collisions with the sample surface, \citeauthor{bonnail2004variable}\cite{bonnail2004variable} proposed a variable structure control design methodology that incorporates Proportional Integral (PI) control. These studies contributed to understanding the properties of the STM control system and its associated challenges.

There is a growing interest in pushing the boundaries of STM performance and examining the existing tradeoffs between stability, speed, robustness, and the ability to capture surface variations precisely on the sample. Although several control strategies have demonstrated promising control and image reconstruction in the STM \cite{popescu2020observer,besanccon2022closed}, commercial STMs still rely on PI controllers because of their simplicity and ease of understanding. Nonetheless, further investigation is needed, and innovative feedback control systems are required to unlock the full potential of the STM and to gain deeper insight into the obtained topography images.

The stability of the closed-loop system under PI control was investigated in \cite{Tajaddodianfa2016, Tajaddodianfa2018} where we demonstrated that variations in the local barrier height (LBH) can result in instabilities when fixed PI gains are employed. Building upon this analysis, we proposed a self-tuning PI controller design that continuously adjusts the PI gains based on LBH measurements to prevent instability \cite{Tajaddodianfa2019}. 

In a separate study \cite{alemansour2021ultrafast}, we implemented a PI controller-based feedback loop that is based on the natural logarithm of the differential conductance, $ln(Rdi/dV)$, instead of the natural logarithm of the tunneling current \cite{oliva1995analysis,voigtlander2015scanning}. This control system can regulate the tip-sample distance, even when the applied sample DC bias voltage is zero. As a result, it provides valuable insights into the interaction of electronic states throughout the entire range of sample bias voltage. An additional advantage is the rapid acquisition of current-voltage (I-V) curves, significantly improving speed and efficiency compared to conventional spectroscopy techniques. In another work \cite{mishra2024kalman}, a Kalman filter-based estimation was discussed around PI control of the STM to simultaneously obtain surface conductivity, $\sigma$, and true topography, $h$, of a sample surface.

The current research introduces a new feedback control mechanism for imaging and lithography applications. In this method, the controller dynamically adjusts the vertical position of the tip while modulating the controller output to maintain a constant value of the gradient of the tunneling current, i.e., $di/dz$, throughout the raster scanning process. By closing the feedback loop on  $ln(Rdi/dz)$, we achieve remarkable improvement in image quality. This innovative imaging mode enables us to acquire highly precise topography images.

An important advantage of this new control method is its ability to measure surface variations accurately. This is achieved by maintaining a constant $ln(Rdi/dz)$ and leveraging the new feedback control loop dynamics throughout the scanning process. This approach significantly improves the sensitivity of the tip/imaging system to subtle changes in surface topography.
The controller output, while feedback is closed on the $ln(Rdi/dz)$, maps the surface to generate images that reveal the topography and structural details of the sample, enabling precise nanoscale imaging and analysis using STM.

In the remainder of this paper, we briefly discuss the theory of tunneling current, the mathematical framework, and the system identification process for the proposed feedback approach in Section\,\ref{sec:modulation}. Subsequently, Section\,\ref{s:Exp_setup} details the experimental setup, presents imaging and lithography results using the new feedback loop, and compares phenomena to conventional STM control method. Finally, conclusions are discussed in Section\,\ref{s:Conclusion4}.

\section{\label{sec:modulation}STM \& Modulation Techniques}

In this section, we briefly explain the tunneling phenomenon in the STM and review conventional modulation techniques used to characterize surface properties. We then introduce a new feedback control approach and showcase our findings, which result in superior topography images.

\subsection{The Tunneling Phenomenon}
\label{s:Tunnel}
The STM obtains images by scanning the tip over a surface. A tungsten (W) probe with a sharp tip is brought near the surface. When the two surfaces are sufficiently close for their wave functions to overlap, typically 0.5-5 \,$nm$, there is a finite probability that electrons will tunnel across the barrier when a bias voltage,$V_b$ (in Volts) is applied between the sample and the tip \cite{hamers1987imaging,binnig1984electron}. The resulting tunneling current, $i$, is approximated by 
\begin{equation} \label{eq1}
i  = f(\sigma, V_b) e^{-1.025 \sqrt{\varphi}\delta }  
\end{equation}

where $\delta =  {z_t} - h$ (in nm) is the sample-tip separation, $z_t$ is the tip displacement due to  controller output, ${u}$, and $h$ is the surface topography. Also, $\varphi$ (in eV or nA/nm) is the average work function of the tip and sample, known as the LBH, and $\sigma$ (in nA/V) represents the electron density of the sample. Note that $\sigma$ and $\varphi$ are surface parameters that define the physical and electronic properties of the surface. Although this equation simplifies the tunneling current model, it effectively explains the operation of the STM. Notably, the current exhibits an exponential dependence on the tip-sample separation: a decrease of $0.1$ \,$nm$ in tip-sample distance results in a two-order magnitude increase in current. This sensitivity allows the tunneling current to control the tip-sample separation precisely with a very high vertical resolution. However, these atomic-resolution images are achievable only under optimal conditions. Larger sample-tip separations or blunt tips tend to blur the localized structures, resulting in topographic images with reduced resolution.

\begin{figure}[h!]
\centering{}
\includegraphics[scale=0.8]{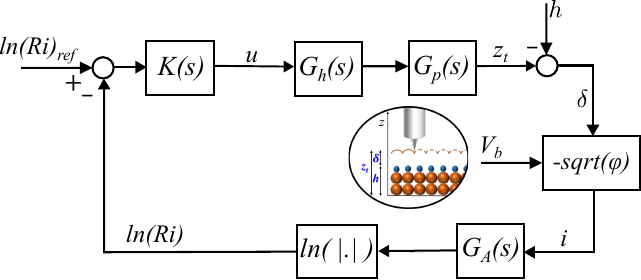}
\caption{Control block diagram of a constant current imaging feedback loop. The controller command, $u$, is amplified by the high voltage amplifier ${G_{h}(s)}$, which drives the piezo-actuator ${G_{p}(s)}$. The tunneling current, $i$, changes momentarily when the tip encounters an unknown surface feature, ${{h}}$. This change is regulated by the controller ${K(s)}$ by adjusting the tip-sample gap ${\delta}$ (where ${\delta}$ is obtained as the tip displacement due to controller output and true sample topography, $z_t - h$) as shown in the image within the circle. The preamplifier ${G_{A}(s)}$ converts the small tunneling current, $i$ to a measurable voltage. Then, the logarithmic amplifier is applied to the absolute value of the signal obtained from the preamplifier as we are working with negative bias voltage.}
\label{f:CC}
\end{figure}

In constant current imaging, a feedback mechanism ensures a steady current by applying a constant bias to the sample, with the tip grounded. As the tip traverses the surface, the constant current feedback regulates the tunneling current and, subsequently, the tip-sample separation, $\delta$. To construct a linear feedback control loop, we take the natural logarithm of the tunneling current in Eq.\,\ref{eq1} to create a linear relationship between $ln (Ri)$ and $\delta$, $i.e.$ 
\begin{equation} \label{eq2}
ln(Ri)  = ln(f(\sigma, V_b)) - 1.025\sqrt{\varphi} \delta  
\end{equation}
where $R$ is the gain of the trans-impedance amplifier (in V/nA) used for converting the small tunneling current to a measurable voltage.
 The control signal required for these vertical adjustments forms the topography image, which depicts a contour of constant charge density on the surface. The control block diagram to perform STM imaging is shown in Fig.\,\ref{f:CC}. Constant current topography also provides insights into the electronic structure of the surface, though distinguishing between electronic and geometric contributions can be challenging. Modulation techniques are employed to extract this spectroscopic information, as detailed and discussed in Section\,\ref{s:LIA}.

\subsection{The Lock-In Amplifier}
\label{s:LIA}

\begin{figure}[h!]
\centering{}
\includegraphics[scale=0.25]{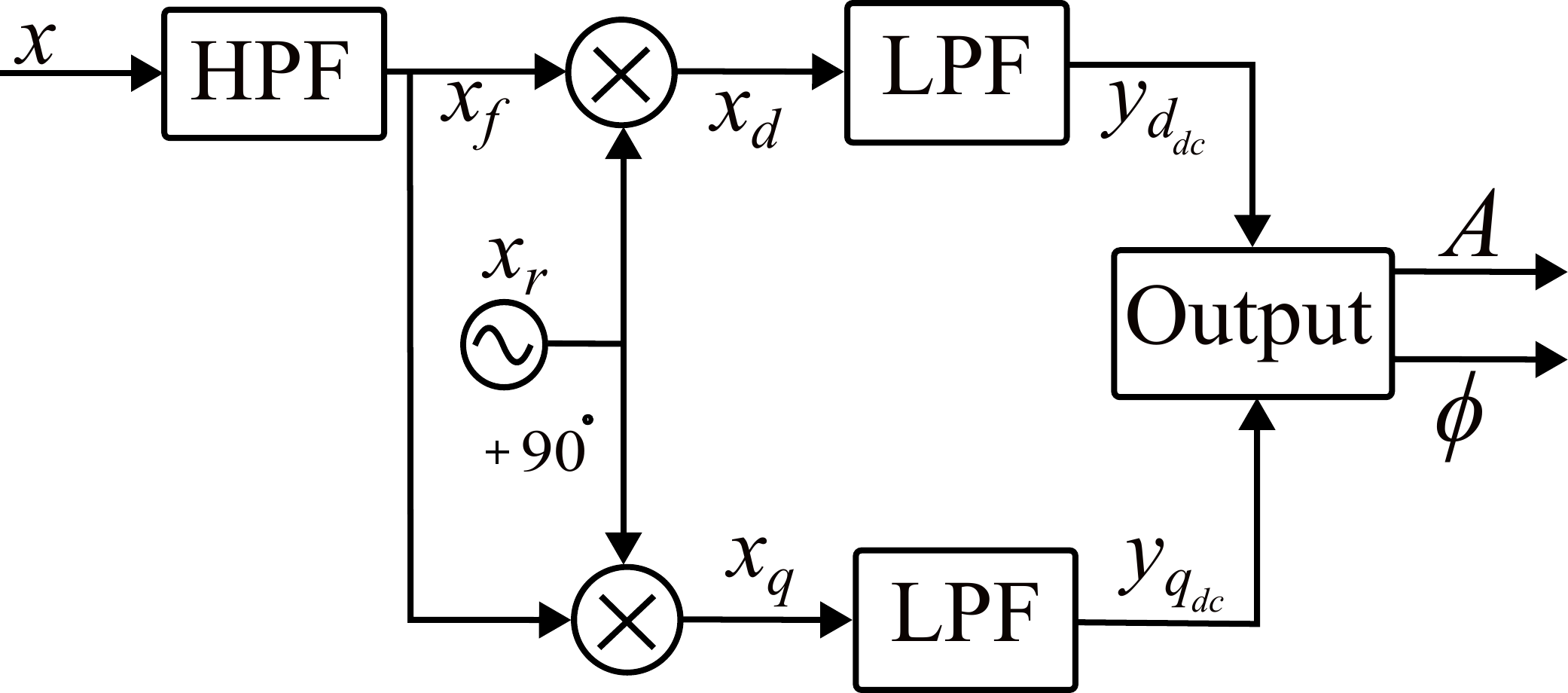}
\caption{Block diagram of a Lock-in amplifier (LIA) implementation. For better readability, the explicit time-dependency of signals is dropped in this block diagram.}
\label{f:LIA}
\end{figure}

Estimation of the surface parameters in scanning tunneling microscopy predominantly relies on utilizing a lock-in amplifier (LIA). Several spectroscopy approaches use LIA owing to its superior demodulation capabilities. The narrow tracking bandwidth of this tool significantly reduces sensitivity to extraneous frequency components and enhances the ability to recover the pertinent signal from noisy measurements accurately. An LIA can also effectively track the amplitude of any relevant frequency component within the input signal, $x(t)$, as shown in  Fig.\,\ref{f:LIA}. 

Based on the LIA and discussion in Appendix \ref{app:subsec}, scanning tunneling spectroscopy (STS) is the most common modulation technique for obtaining various surface parameters. In this method, bias voltage, $V_b$, is modulated and tunneling current, $i$ is obtained as $\mathit{f} (V_b + V_m sin(\omega t))$. This tunneling current, when passed through LIA, tuned at $\omega_r = \omega$, gives
\begin{equation} \label{eq8}
  \Bigl(\frac{\partial\;ln\;i}{\partial\;ln\;V}\Bigr) = \sigma
\end{equation}
In the simplest approximation, for a small amplitude of modulation voltage, the first derivative of the tunneling current model equation is proportional to the local density of states (LDOS) or surface conductivity of the sample \cite{feenstra1994scanning,alemansour2021high,hamers1989atomic}. 

The gap modulation method, initially introduced as an early STM spectroscopy technique, is employed to measure the local barrier height, $\varphi$ \cite{jia1998variation,maeda2004local,lang1988apparent,wiesendanger1987local}. This method involves applying a modulating signal at frequency $\omega_r = \omega$ to the controller output, ${u}$, inducing tiny tip oscillations in a direction normal to the sample surface.  Consequently, tracking the amplitude of $ln(Ri)$ at frequency $\omega_r = \omega$ using a lock-in amplifier to obtain $\varphi$ as follows.
\begin{equation} \label{eq9}
 \Bigl( \frac{\partial\;ln\;i}{\partial z}\Bigr)^2 \approx \varphi
\end{equation}

Surface parameters obtained from Eq.\,\ref{eq8} and \,\ref{eq9} are influenced by the local electronic properties of both the surface and the tip. Therefore, they are called the local density of states (LDOS) and the LBH, respectively. These measurements serve as the foundation for creating surface conductivity and LBH images, which provide additional insights into surface physics, revealing more detailed information about the surface. The information provided by these image data, as shown in  Fig.\ref{f:LBH_Sigma}, helps us establish the effectiveness of the proposed method in Section\,\ref{s:newM}.

\subsection{The proposed feedback loop}
\label{s:newM}

Here, we propose a novel STM feedback control method that builds on our recent research efforts that involve closing the feedback loop on differential conductance, ${di/dV}$ measurements, as detailed in our prior publications \cite{alemansour2021ultrafast, moheimani2023methods,alemansour2021high}. That work demonstrated the potential benefits of utilizing ${di/dV}$ feedback to obtain a symmetric $I-V$ curve for the sample surface, among other things. 

\begin{figure}[h!]
   \centering
    \includegraphics[scale = 0.46]{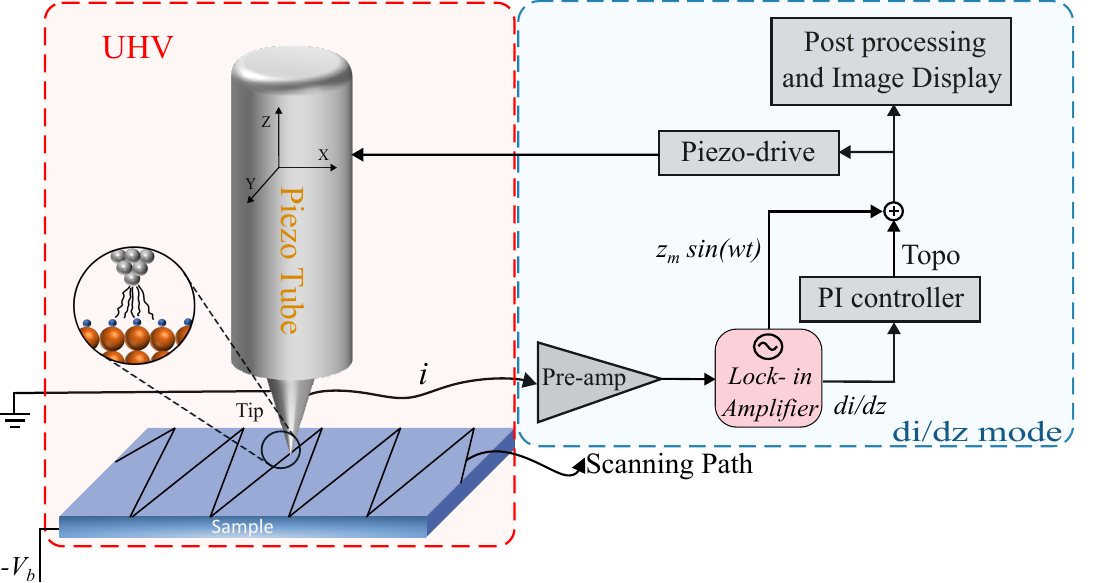}
    \caption{Schematics of the scanning tunneling microscope working in the constant $di/dz$ mode.}
  \label{f:sch_didzorg}
\end{figure}

Building upon this concept, we propose a feedback control mechanism designed explicitly for imaging and lithography, including in future platforms that utilize multiple STM tips for high-throughput lithography\cite{alipour2021mems,alipour2024atom}. Our method leverages the insights gained from our earlier studies to understand the performance of the new feedback loop. This new approach redefines the STM feedback control strategy and broadens the STM's applicability to more demanding and intricate operations in nanoscale imaging and fabrication. The schematic of the proposed feedback implementation on $ln(Rdi/dz)$ is shown in Fig.\ref{f:sch_didzorg}.

\begin{figure}[h!]
   \centering
    \includegraphics[scale = 0.65]{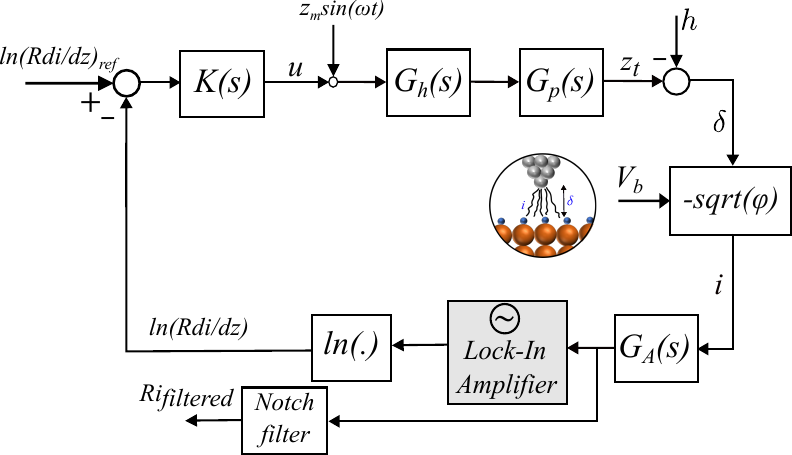}
    \caption{The control block diagram of the z-axis of STM in the constant ${di/dz}$ mode. In this mode, a modulation signal is applied to the controller output, $u$. The controller command is amplified by the high voltage amplifier ${G_{h}(s)}$, which drives the piezo-actuator ${G_{p}(s)}$. The ${di/dz}$ changes momentarily when the tip encounters an unknown surface feature, ${{h}}$. This change is regulated by controller ${K(s)}$ by adjusting the tip-sample gap ${\delta}$ (where ${\delta}$ is obtained as the tip displacement due to controller output and true sample topography, $z_t - h$). The preamplifier ${G_{A}(s)}$ converts very small tunneling current to a measurable voltage. The amplified current is passed through LIA to obtain ${di/dz}$. The positive valued signal obtained from the LIA is then applied to a logarithmic.}
  \label{f:didzorg}
\end{figure}

In this method, the tip-sample distance is controlled by a feedback controller that regulates the ${di/dz}$, i.e., by keeping ${ln(Rdi/dz)}$ constant throughout a scan. The concept is depicted in Fig.\ref{f:didzorg}.

From Eq.\,\ref{eq1}, we may write
\begin{equation}\label{10}
    \frac{di}{dz} = -1.025\sqrt{\varphi}f(\sigma, V_b) e^{-1.025 \sqrt{\varphi}\delta }
\end{equation}
 The modulation technique can simultaneously acquire the ${di/dz}$ signal with the topography image. Experimentally, this is obtained by superimposing a sinusoidal modulation signal, $z_m$ sin$(\omega t)$ to the controller output, $u$, and measuring the amplitude of the AC component of tunneling current at the modulating frequency by an LIA. In this mode, the tip is scanned over the surface while the feedback loop keeps the tunneling current amplitude at modulating frequency constant, as illustrated in Fig.\ref{f:didzorg}. 

\begin{figure*}[t!]
 \includegraphics[scale=0.32]{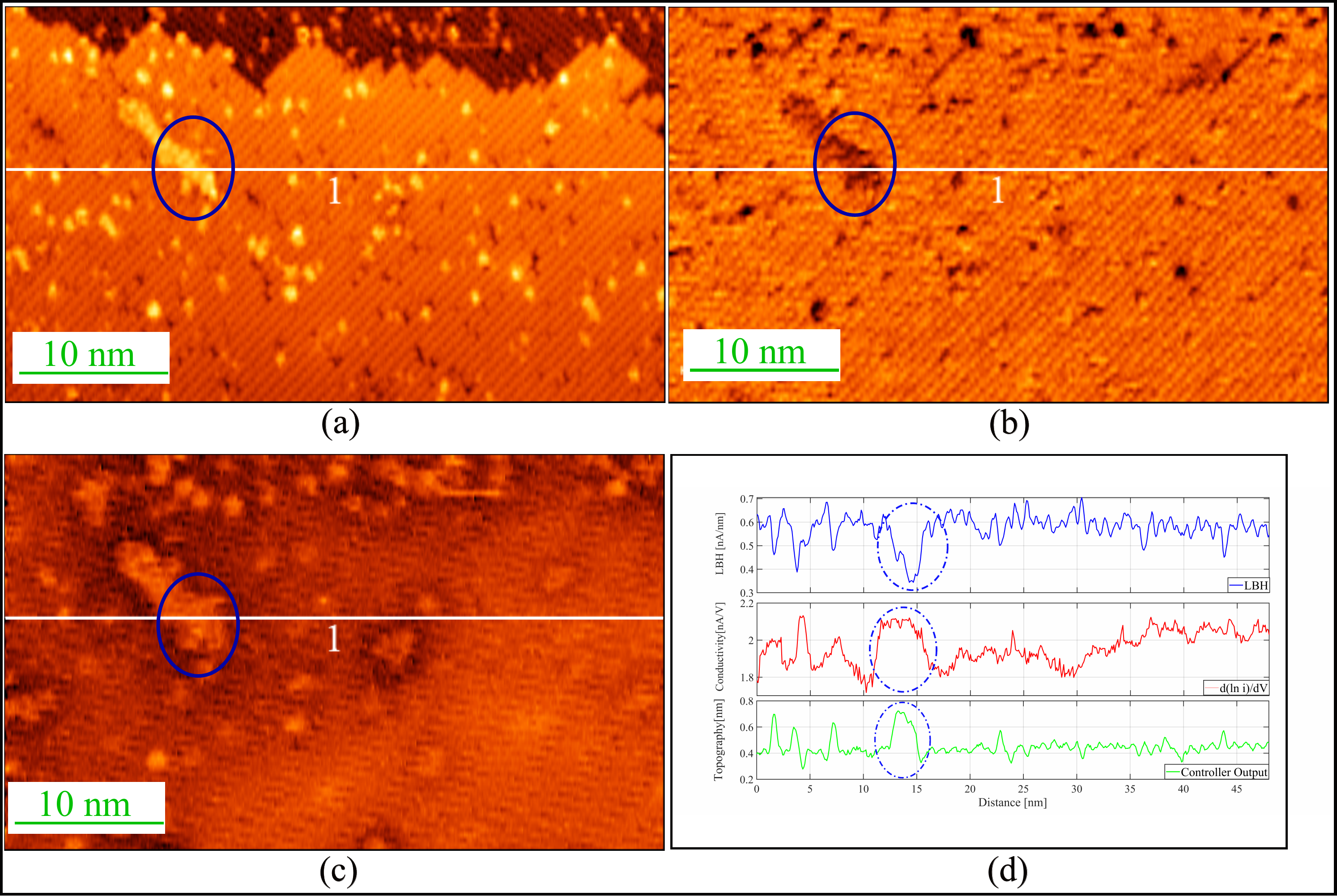}
 \caption{The (a) Topography image, (b) LBH image, and (c) Conductivity image of a $\mathrm{Si(100)-2\times 1:H}$  passivated surface ($\sigma$ obtained from Eq.\,\ref{17}). (d) The profile of line 1 on all three images is represented together. The sample surface highlighted with a blue circle shows dangling bonds produced through hydrogen depassivation lithography (HDL), highlighting the change in surface properties. Imaging and lithography are performed with a constant current feedback loop. All the images are obtained simultaneously.}
 \label{f:LBH_Sigma}
\end{figure*}

This amounts to a new imaging mode for the ultra-high vacuum (UHV) STM. As shown in the corresponding control system block diagram in Fig.\ref{f:didzorg}. We close the STM feedback loop on the $ln({Rdi/dz})$ signal. The set-point minus the natural logarithm of  ${di/dz}$ signal is defined as the error. The proportional integral controller regulates the $di/dz$ and, thus, the tip-sample distance to minimize the error signal. The controller output is then plotted along with the X and Y positions of the tip to construct a topography image of the surface.
  
To understand the underlying principle governing this feedback control loop, we need to first linearize equation Eq.\,\ref{10}, i.e.
\begin{equation}\label{11}
    \ln\left(R\frac{di}{dz}\right) = \ln((-1.025\sqrt{\varphi}\;f(\sigma, V_b)))-1.025 \sqrt{\varphi}\delta
\end{equation}

We can make two immediate observations. First, we learn from Eq.\,\ref{11} that $\ln(Rdi/dz)$ is a linear function of $\delta$. Thus, by regulating $\ln(Rdi/dz)$, we regulate $\delta$, all other parameters being constant. To understand the second observation, we point out that all surface properties and disturbances are captured in the controller output, i.e., the control signal contains information about all of these disturbances. In the STM feedback control system, there are two disturbances that are of fundamental importance: one is the surface variation, denoted as $h$, and the other is the output disturbance represented by ${\ln(f(\sigma, V_b))}$ in the constant-current feedback loop\cite{mishra2022kalman, mishra2024kalman}. It is important to note that while $h$ represents the surface topography, ${\ln(f(\sigma, V_b))}$ captures the effect of surface electronic properties, i.e., electronic disturbances.

In the constant $\ln(Rdi/dz)$ feedback loop, we observe that $\ln((-1.025\sqrt{\varphi}\,f(\sigma, V_b)))$ acts as an output disturbance. Referring to Fig.\,\ref{f:LBH_Sigma}, which illustrates data obtained from a $\mathrm{Si(100)-2\times 1:H}$ surface, we note that the LBH is smaller when the surface conductivity $\sigma$ is larger, and vice versa. Therefore, surface conductivity and barrier height exhibit opposite characteristics for a $\mathrm{Si(100)-2\times 1:H}$ surface. Thus, the effect of the output (electronic) disturbance on the new STM feedback loop moderates as the tip moves from one point to another during imaging. For both feedback loops, the input disturbance, i.e., surface variations, $h$, remains the same. However, each feedback loop is subjected to a different output disturbance since surface electronic properties affect each loop differently. In the conventional current feedback control, the output disturbance could undergo more substantial variations as the tip moves from one location on the surface to another, making it challenging for the controller to obtain the same level of performance compared to the $\ln(Rdi/dz)$ feedback case. Hence, the tip has to come closer to the surface to obtain high-resolution images in the constant current imaging mode compared to the constant $di/dz$ imaging mode. The same applies when the tip operates in the HDL mode under $di/dz$ feedback.

Fig.\,\ref{f:LBH_Sigma} shows that the LBH is almost two times lower over the dangling bonds (bright contrast over a profile). This change implies an increase in the gain of the STM system by nearly a factor of two \cite{Tajaddodianfa2019}. This change in LBH appears in the first term in Eq.\,\ref{11} and helps reduce the effect of output disturbance on the new feedback loop on $\ln(Rdi/dz)$ and hence could be contributing to the stability of the new feedback loop and therefore tip images with better resolution.

\subsection{Model discovery for the $ln(Rdi/dz)$ feedback loop}
\label{s:sysID}
\begin{figure*}[t!]
\centering{}
\includegraphics[scale=0.8]{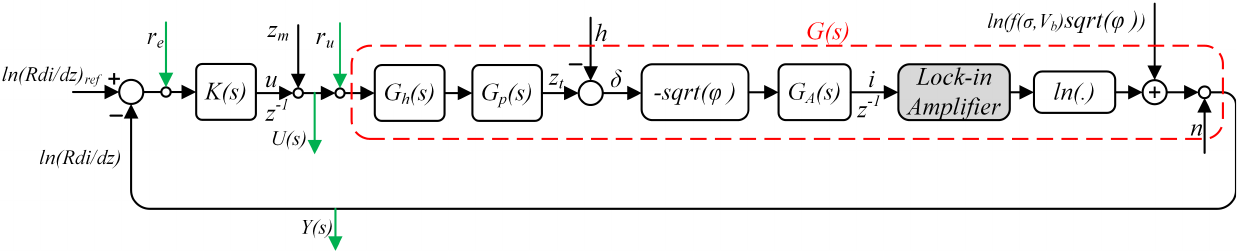}
\caption{System identification of z-axis control system in the constant $ln(Rdi/dz)$ mode.}
\label{f:SysId}
\end{figure*}

To design a controller for this feedback loop, acquiring a mathematical model of the open loop system, $G(s)$ is essential. To achieve this, we perform system identification tests in the frequency domain by utilizing the linearized system depicted in Fig.\,\ref{f:SysId}. These tests are conducted to extract the transfer function of $G(s)$ from the controller output, $u$ to the STM output, $ln(Rdi/dz)$. The system identification experiment must be performed with the feedback loop enabled and tunneling current established \cite{Tajaddodianfa2016}.

\begin{figure*}[t!]
\centering{}
\includegraphics[scale=0.35]{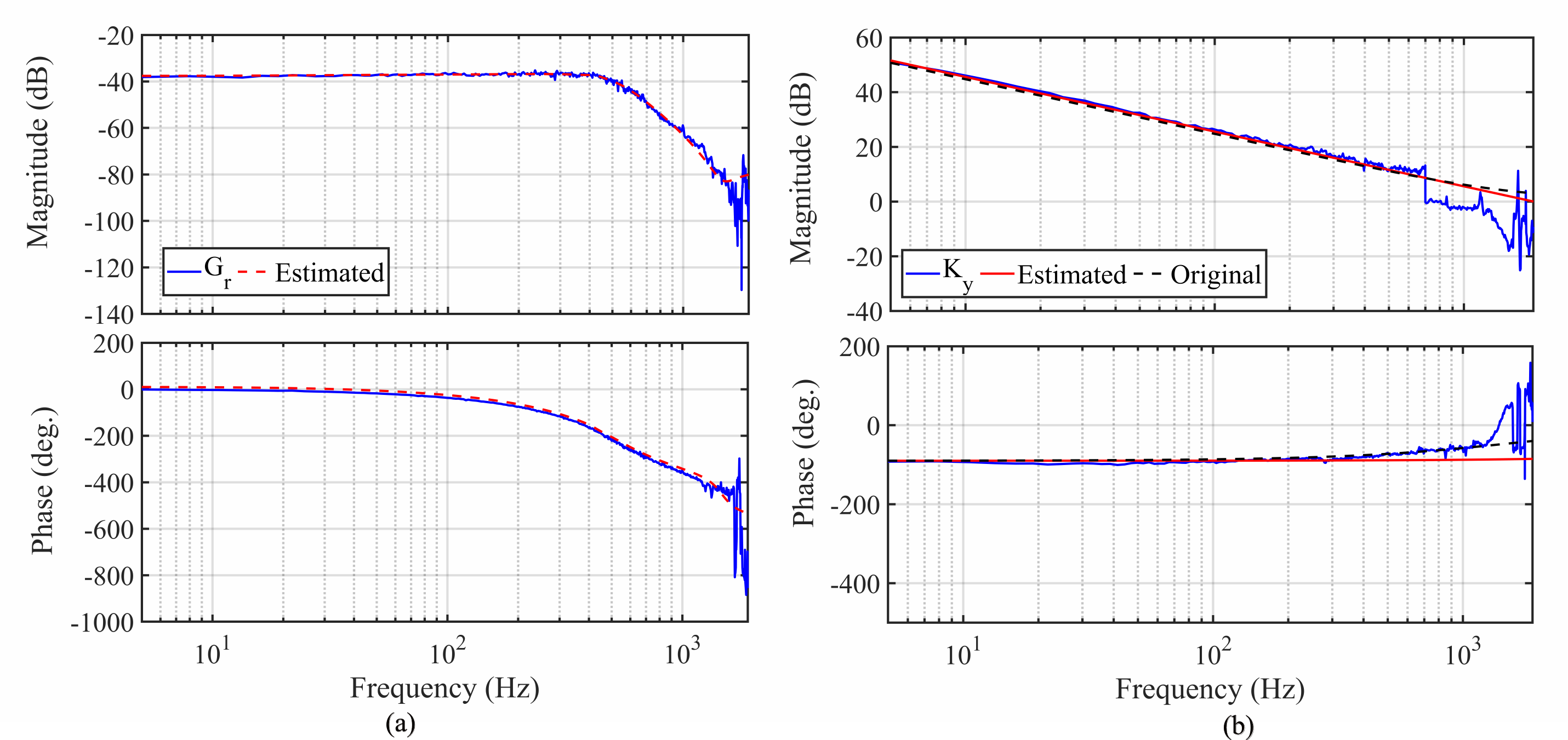}
\caption{Measured and estimated open-loop FRFs. (a) The STM plant, $G(s)$, and (b) the PI controller, $K(s)$ (Eq.\,\ref{16} and Eq.\,\ref{17}). The PI controller is set with $k_i$ = 1.625$\times$ $10^4$ and $\omega_c$ = $10^{4}\, rad/sec$. This is the system identification of a home-built Lyding scanner.}
\label{f:BodeEst}
\end{figure*}
We adopt the methodology developed in \cite{forssell1999closed, Tajaddodianfa2016} for measuring the frequency response of $G(s)$. We add exogenous signals $r_e$ and $r_u$ at the set-point and controller output, $u$, and simultaneously record the responses at the input, $U(s)$, and output, $Y(s)$, of the plant, as shown in Fig.\,\ref{f:SysId}. By varying the frequency of these exogenous signals from 5\,$Hz$ - 1500\,$Hz$, we record the output signals $Y(s)$ and $U(s)$, which correspond to $ln(Rdi/dz)$ and $u$, respectively. We perform multiple measurements and average the outputs at each frequency point to reduce the effect of measurement noise. We add each exogenous signal at a time, obtain a frequency response, and repeat the experiment twice to obtain complete plant and controller dynamics information. The following transfer functions between the exogenous inputs and outputs can be derived based on the data and the configuration in Fig.\,\ref{f:SysId}. 

\begin{equation} \label{12}
	 	{G_{r_{e}U}(s)} = \frac{K(s)}{1+ K(s)G(s)}
	\end{equation}
 \begin{equation} \label{13}
	 	{G_{r_{e}Y}(s)} = \frac{K(s)G(s)}{1+ K(s)G(s)}
	\end{equation}
  \begin{equation} \label{14}
	 	{G_{r_{u}U}(s)} = \frac{-K(s)G(s)}{1+ K(s)G(s)}
	\end{equation}
  \begin{equation} \label{15}
	 	{G_{r_{u}Y}(s)} = \frac{G(s)}{1+ K(s)G(s)}
	\end{equation}

A frequency response of the STM transfer function $G(s)$ is obtained by dividing the frequency response functions (FRFs) at each frequency point as shown in  Fig.\,\ref{f:BodeEst} a. A $7^{th}$ order model was then fitted to the measured frequency response $G(j\omega)$ obtained from Eq.\,\ref{16}
 \begin{equation} \label{16}
	 	{G(j\omega)} = {G_r(j\omega)} = \frac{G_{r_{e}Y}(j\omega)}{G_{r_{e}U}(j\omega)} 
	\end{equation}
 Similarly, we can obtain the dynamics of the actual controller as shown in Fig.\,\ref{f:BodeEst} b, by fitting a model to measured $K(j\omega)$. 
 \begin{equation} \label{17}
	 	{K(j\omega)} = {K_y(j\omega)} = \frac{G_{r_{e}Y}(j\omega)}{G_{r_{u}Y}(j\omega)} 
	\end{equation}
Conventionally, a PI controller is employed in STM closed-loop control systems to regulate the system close to the set-point. The PI controller is mathematically described as:
\begin{equation} \label{17}
	 K(s) = k_i \bigg(\frac{1}{s} +\frac{1}{\omega_{c}}\bigg)
\end{equation}
where $k_i$ is the overall gain and $\omega_{c}$ is the corner frequency.

\begin{figure}[h!]
\centering{}
\includegraphics[scale=0.03]{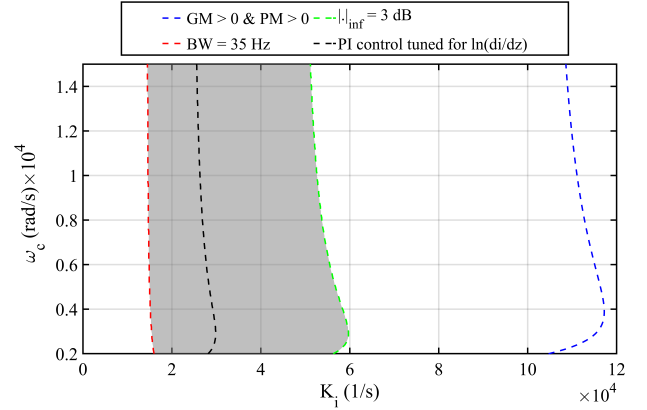}
\caption{Stable PI control values within grey region for the constant $ln(Rdi/dz)$ imaging.}
\label{f:Stability}
\end{figure}

 The parameters of the PI controller are crucial in determining the performance and stability of the closed-loop system. Having identified a model for $G(s)$, (Eq.\,\ref{16}), we may proceed to design an effective PI controller analytically. According to~\cite{Tajaddodianfa2019}, three key criteria are recommended for the design of the STM-PI controller. The permissible range of integrator gains $k_i$ can be determined for each frequency $\omega_{c}$ based on specified criteria, as illustrated in Fig.\,\ref{f:Stability} for a specific STM system. The stability margin criterion dictates the maximum allowable values for $k_i$. Thus, integrator gain follows 0 < $k_i$ < Gain Margin (GM), depicted by the blue curve in Fig.\,\ref{f:Stability}. Conversely, we have chosen the minimum imaging bandwidth of 35\,$Hz$, imposing a lower limit on $k_i$, represented by the red curve. As a standard rule ~\cite{phillips2007digital,teo2018comparison}, the imaging bandwidth is 10-100 times greater than the scan frequency, which in turn is guided by the mechanical dynamics of the system. Additionally, the infinity norm of the imaging transfer function must stay below a pre-defined threshold of 3\,$dB$ ~\cite{ Tajaddodianfa2019}. To satisfy these criteria, the PI controller parameters should be selected from the highlighted grey area in Fig.\,\ref{f:Stability}. The black dashed line indicates the recommended integrator gain, set at half the maximum permissible threshold of 3\,$dB$. 
Based on the above understanding, manual tuning of the STM PI gains involves first selecting an appropriate $\omega_{c}$ value. This $\omega_{c}$ is selected to sufficiently mitigate the nonlinear effects. Therefore, $\omega_{c} =10^{4}\, rad/sec$ is chosen, as it effectively reduces nonlinear effects and remains below the first resonant peak of the system, as shown in  Fig.\,\ref{f:BodeEst}. Additionally, a low-pass filter is in the feedback loop, which attenuates high-frequency noise and reduces the amplitude of the first resonant peak. This design choice allows the cutoff frequency to be set near the first resonant peak without compromising stability. Next, the $k_i$ gain is gradually increased until ringing is observed in the current signal. The $k_i$ gain is then adjusted to half the value at which the ringing was observed, setting this as the operational value.

\begin{figure}[h!]
\includegraphics[scale=0.03]{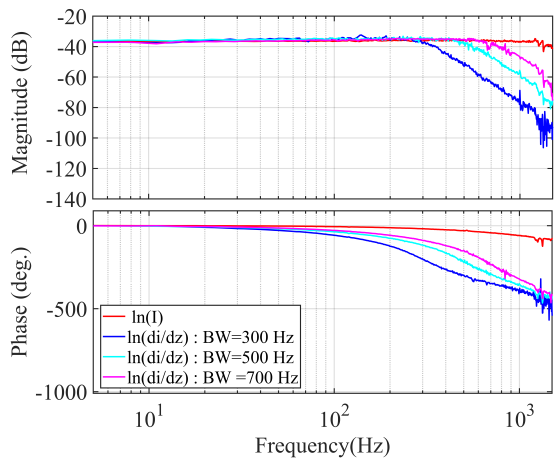}
\caption{Frequency responses of the STM system operating in constant $ln(Rdi/dz)$ mode. A modulation voltage of 2\, $kHz$ with an amplitude of 0.8\, $mV$ was superimposed on the controller output, $u$. The resultant current was demodulated at the fundamental frequency. The feedback loop was engaged with the amplitude of the demodulated signal, maintaining a set-point value of 0.5\,$nA/nm$. Frequency responses were recorded for LIA LPF cutoff frequencies set to 300\,$Hz$, 500\,$Hz$, and 700\,$Hz$. For comparative purposes, the system's response in constant current imaging mode was also plotted.}
\label{f:bodeLIA}
\end{figure}

 The bandwidth of the open-loop system $G(s)$ is primarily determined by the low-pass characteristics of the LIA, particularly the settings of the LPF. Fig.\,\ref{f:bodeLIA} presents the open-loop frequency responses corresponding to three distinct LPF cutoff frequencies. The results demonstrate that the open-loop system's bandwidth increases proportionally with the LPF cutoff frequency. The primary function of the low-pass filter is to eliminate out-of-bandwidth noise and disturbances, thus improving the signal-to-noise ratio (SNR) of the demodulated signals. However, a narrower LPF bandwidth, which enhances the SNR, necessitates operating the system at a reduced scan speed. This trade-off provides a unique opportunity to reject high-frequency noise, which may excite the resonant frequency while scanning at the required bandwidth and avoiding tip crashes.

\section{Experimental Results}
\label{s:Exp_setup}

In this section, we present our experimental findings. Firstly, we detail the experimental setup, presenting the results highlighting the difference between the conventional and the constant ${di/dz}$ STM modes. We perform a series of imaging and lithography experiments to demonstrate the performance of the new feedback control system for comparison with the conventional approach.

\subsection{Experimental setup}
\label{s:Exp_results}

We conducted several experiments with a home-built Lyding scanner-based STM and a ScientaOmicron Variable Temperature Scanning Probe Microscope (VT SPM). Both scanners are maintained at room temperature in a UHV environment of $10^{-11}$\,Torr. The sample used for imaging is 4 $\times$ 10$\;mm^{2}$, H-passivated, Si\,(100)-2$\times$1 surface prepared under UHV conditions~\cite{lyding1994nanoscale}. The sample is degassed at 650$^\circ$C for 8 - 10$\; hrs$ and flashed 5 times to 1240$^\circ$C for 30$\;sec$. The sample is cooled to  300$^\circ$C and is then exposed to a flux of H atoms for 10$\;min$ at $10^{-6}$\; Torr. A monolayer of H atoms forms on the sample surface that functions as a resist. The atomic H is generated by a 1300$^\circ$C Tungsten filament cracking the background $H_2$ molecules into atomic H \cite{walsh2009atomic}. 

Current measurements on the home-built STM are performed by a Femto DLPCA-200 current pre-amplifier, which features a gain of $1\; V/nA$ and a $1.1\;kHz$ bandwidth. For ScientaOmicron VT SPM, we use a preamplifier setting of $330\; nA$, with a gain of $0.03\; V/nA$ and a bandwidth of $3\;kHz$ or $100\;kHz$. For ScientaOmicron VT SPM, we use an unfiltered current setting. The system is controlled by a 20-bit Digital Signal Processing (DSP) unit, operating at $100 \;kHz$ sampling rate, supporting imaging and lithography operations. This DSP unit features D/A and A/D channels with a 20-bit resolution at $\pm10 \; V$. This system also includes a low-noise, high-bandwidth high voltage amplifier with a gain of $13.5$ to provide the required actuation signals to the piezotube. The software and the DSP are known as Scanz\textsuperscript{\texttrademark} and Zyvector\textsuperscript{\texttrademark} \footnote{Scanz\textsuperscript{\texttrademark} and Zyvector\textsuperscript{\texttrademark} are a trademarked product of Zyvex Labs} respectively. The DSP integrates the proposed feedback and allows for adjusting parameters through Scanz\textsuperscript{\texttrademark} via Python programming. The STM images included in this paper were captured using the above setup.

\subsection{The constant $di/dz$ feedback loop}
\label{s:didz}
\begin{figure}[h!]
\centering{}
\includegraphics[scale=0.02]{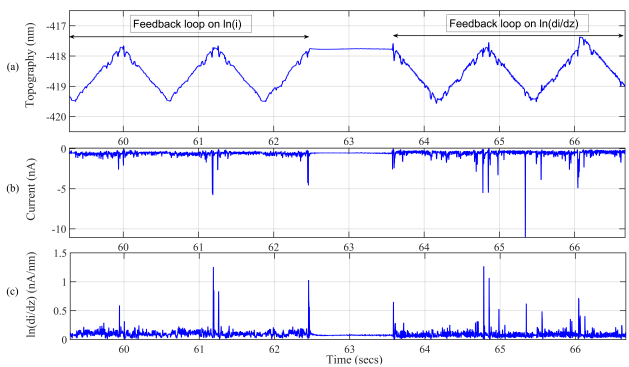}
\caption{Real time data captured to show switchover from $ln(Ri)$ to $ln(Rdi/dz)$ feedback loop. (a) Topography, (b) Tunneling current, and (c) ln($Rdi/dz$). The scanning parameters are as follows: (a) set-point for $ln(Ri)$ = 0.5\,$nA$, scan speed = 100\,$nm/sec$ (b) Obtained $ln(Rdi/dz)$ from LIA and set-point for new feedback controller = 0.165\,$nA/nm$. The user can change to a new set-point once the feedback loop on $ln(Rdi/dz)$ is established. }
\label{f:switchover}
\end{figure}

Using the slip/stick coarse-positioning mechanism, the tip is brought close to the sample so that the sample surface gets within the range of the fine positioner. This procedure is an automated, high-sensitivity stepping process. During this step, the piezoelectric tube extends to its maximum range while the coarse positioner decreases the tip-sample gap. The control system continuously monitors the tunneling current. Once the tunneling current is detected, the piezoelectric tube retracts immediately. This process is repeated till the fine positioner is within its desired extension range. The tunneling current is established at the desired set-point at this stage due to the constant $ln(Ri)$ feedback loop. We start to scan the surface to observe if the tip is stable. We add a sinusoidal modulation signal to the controller output, $u$, to switch to the constant $di/dz$ feedback loop. The amplitude and frequency of the sinusoidal modulation signal are $z_m$ = 0.8\, $mV$ and $\omega$ = \,2\, $kHz$. The modulation amplitude is chosen to limit the tip oscillations to 0.1\,$nm$. The frequency of modulation voltage is selected outside the bandwidth of the closed-loop system. It should also be relatively low so as not to excite the resonant frequency of the system. We use the values mentioned above of the amplitude and modulation frequency of sinusoidal signals for all the experimental results. Adding a modulation signal to the controller output results in high-frequency components in the current signal at harmonics of the fundamental frequency of the modulation signal, as can be obtained from Eq.\,\ref{eq7}. Therefore, we implement five notch filters tuned to the fundamental frequency and all of its higher order multiples to obtain filtered current, as seen in  Fig.\,\ref{f:didzorg}. At this point, we obtain the natural log of $di/dz$ as the output of LIA, as illustrated in Fig.\,\ref{f:switchover}. We then switch to a new feedback controller at $62.5\; sec$. It takes $1\; sec$ to set the new reference value based on $ln(Rdi/dz)$ value as shown in  Fig.\,\ref{f:switchover}.

\subsection{Constant ${di/dz}$ Imaging}
\label{s:Imaging_results}

\begin{figure}[h!]
\includegraphics[scale=0.15]{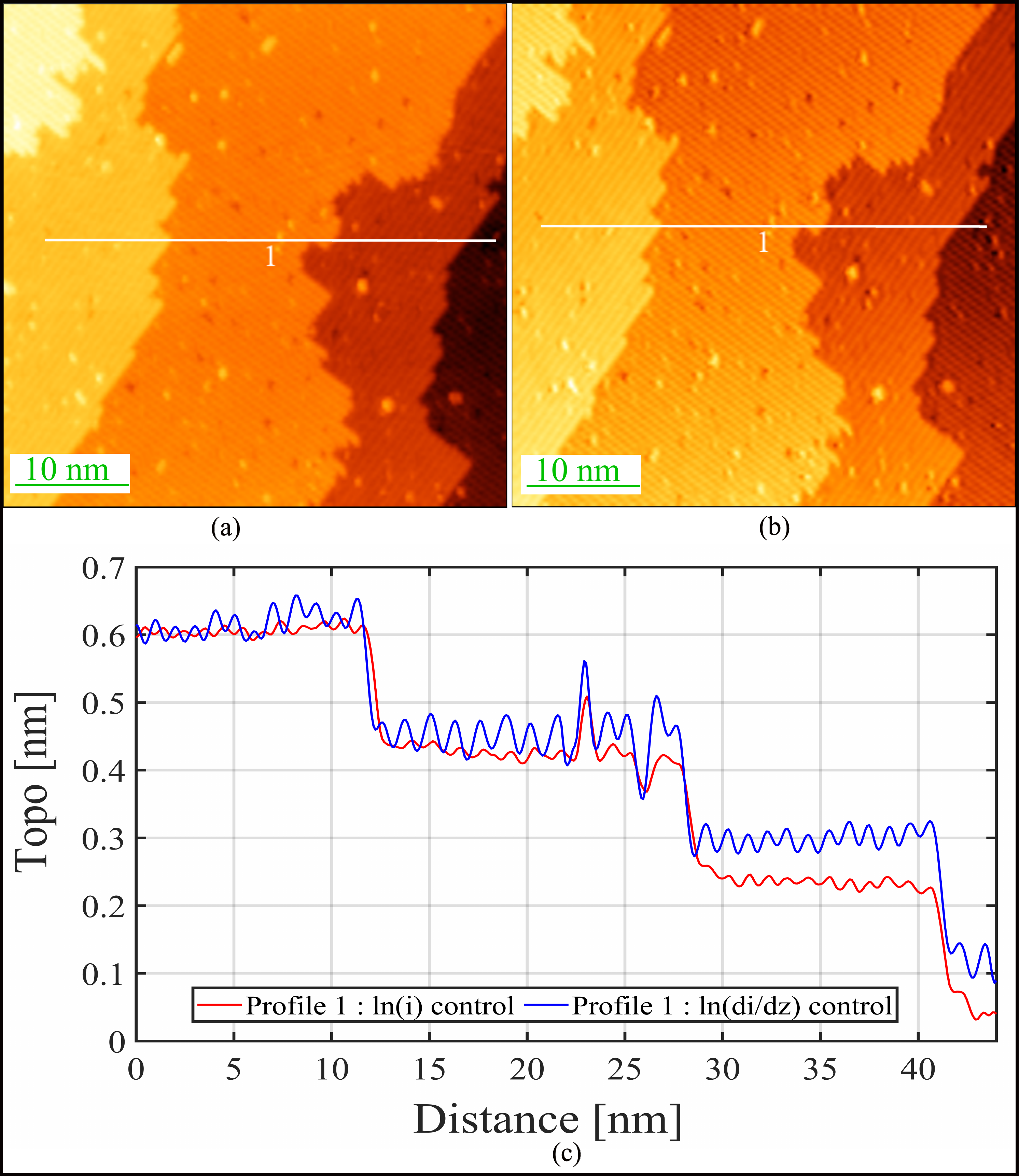}
\caption{The STM images obtained with the constant $ln(Ri)$ and $ln(Rdi/dz)$ feedback loops. Profile 1 is drawn on both images to highlight the observations. (a) Topography image obtained from $ln(Ri)$ feedback loop. (b) Topography image obtained from $ln(Rdi/dz)$ feedback loop. (c) Topography profiles on line 1. The imaging parameters are as follows: set-point for constant $ln(Ri)$ = 0.5\,$nA$, the set-point for constant $ln(Rdi/dz)$ = 0.25\,$nA/nm$.  For both images: bias voltage = -2.5\,$V$, scan size = 48\,$nm$ $\times$ 48\,$nm$, scan speed = 100\,$nm/sec$, image resolution = 512 $\times$ 512\,$pixels$. This image is obtained with a home-built Lyding scanner-based STM.}
\label{f:Figure8}
\end{figure}

\begin{figure*}[t!]
\includegraphics[scale=0.18]{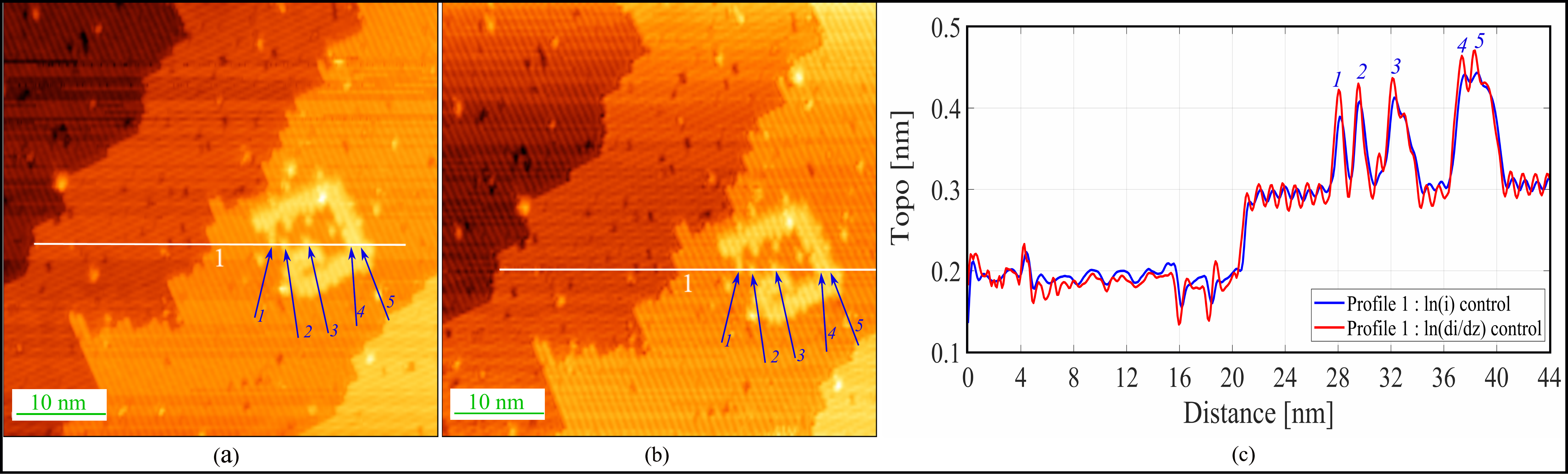}
\caption{The STM images obtained with the constant $ln(Ri)$ and $ln(Rdi/dz)$ feedback loops. Profile 1 was drawn on both images to highlight the observations. (a) Topography image obtained from $ln(Ri)$ feedback loop. (b) Topography image obtained from $ln(Rdi/dz)$ feedback loop. (c) Topography profiles on line 1. The imaging parameters are as follows: set-point for constant $ln(Ri)$ = 0.5\,$nA$, set-point for constant $ln(Rdi/dz)$ = 0.04\,$nA/nm$. For both images: bias voltage = -2.5\,$V$, scan size= 48\,$nm$ $\times$ 48\,$nm$, scan speed = 80\,$nm/sec$, image resolution = 512 $\times$ 512\,$pixels$. Five arrows are marked to highlight the differences in both images. The set-point for constant $di/dz$ loop is set according to the discussion in Section \,\ref{s:didz}. The $ln(Ri)$ feedback loop performs spiral lithography with one loop. These images are obtained with a home-built Lyding scanner-based STM.}
\label{f:Figure10}
\end{figure*}
Once the feedback loop is closed on $ln(Rdi/dz)$, the piezotube rasters the Tungsten (W) tip across the X-Y plane, and the topography of the sample is constructed based on the control commands driving the piezoactuator. We experimented with different tips, samples, and STM scanners. The image in Fig.\,\ref{f:Figure8} b, obtained from the constant $ln(Rdi/dz)$ loop distinctly resolves the rows of dimers on the $\mathrm{Si(100)-2\times 1:H}$ surface, particularly at the step edges in the $S_b$ direction, where the dimer rows are perpendicular to the step edge, as compared to the image obtained from the feedback control loop closed on $ln(Ri)$, shown in Fig.\,\ref{f:Figure8} a. Additionally, the overall contrast obtained in Fig.\,\ref{f:Figure8} b as compared to Fig.\,\ref{f:Figure8} a is highlighted by going over profile 1 on the same sample area. We observe in Fig.\,\ref{f:Figure8} c that the individual dimers appear with better contrast and height than the dimer resolution obtained by constant $ln(Ri)$. The set-point for the new feedback controller is set according to the discussion in Section\,\ref{s:didz}.

The subsequent experimental results were obtained when the tip exhibited mild instability or frequent changes. This scenario provided an opportunity to evaluate the stability and sensitivity of the proposed new feedback loop toward surface features. We conducted spiral lithography with $ln(Ri)$ feedback loop to create dangling bonds and compared the observations from both imaging techniques. Fig.\,\ref{f:Figure10} a and b display topography images acquired with the feedback loop closed on $ln(Ri)$ and $ln(Rdi/dz)$, respectively. A notable observation is the enhanced contrast in Fig.\,\ref{f:Figure10} b compared to Fig.\,\ref{f:Figure10} a. Additionally, the dimer row resolution in both directions is more clearly visible in Fig.\,\ref{f:Figure10} b, showcasing all surface features despite similar tip behavior in both images. To underscore the superiority of the new control method, we draw a profile over dangling bonds and highlight surface features with five arrows. For arrows marked $1$, $2$, and $3$, surface features (dangling bonds) are evident in both Fig.\,\ref{f:Figure10} a and b, but more distinctly in Fig.\,\ref{f:Figure10} b. Arrows marked $4$ and $5$ point to dangling bond formations along the dimer rows. In Fig.\,\ref{f:Figure10} a rows of dangling bonds appear as one broad bright contrast, whereas Fig.\,\ref{f:Figure10} b distinctly reveals three dimer rows. This detail is further highlighted in Fig.\,\ref{f:Figure10} c. The superiority of the new control method is evident in the greater detail of surface features observed despite tip instability.

\begin{figure}[h!]
\centering{}
\includegraphics[scale=0.15]{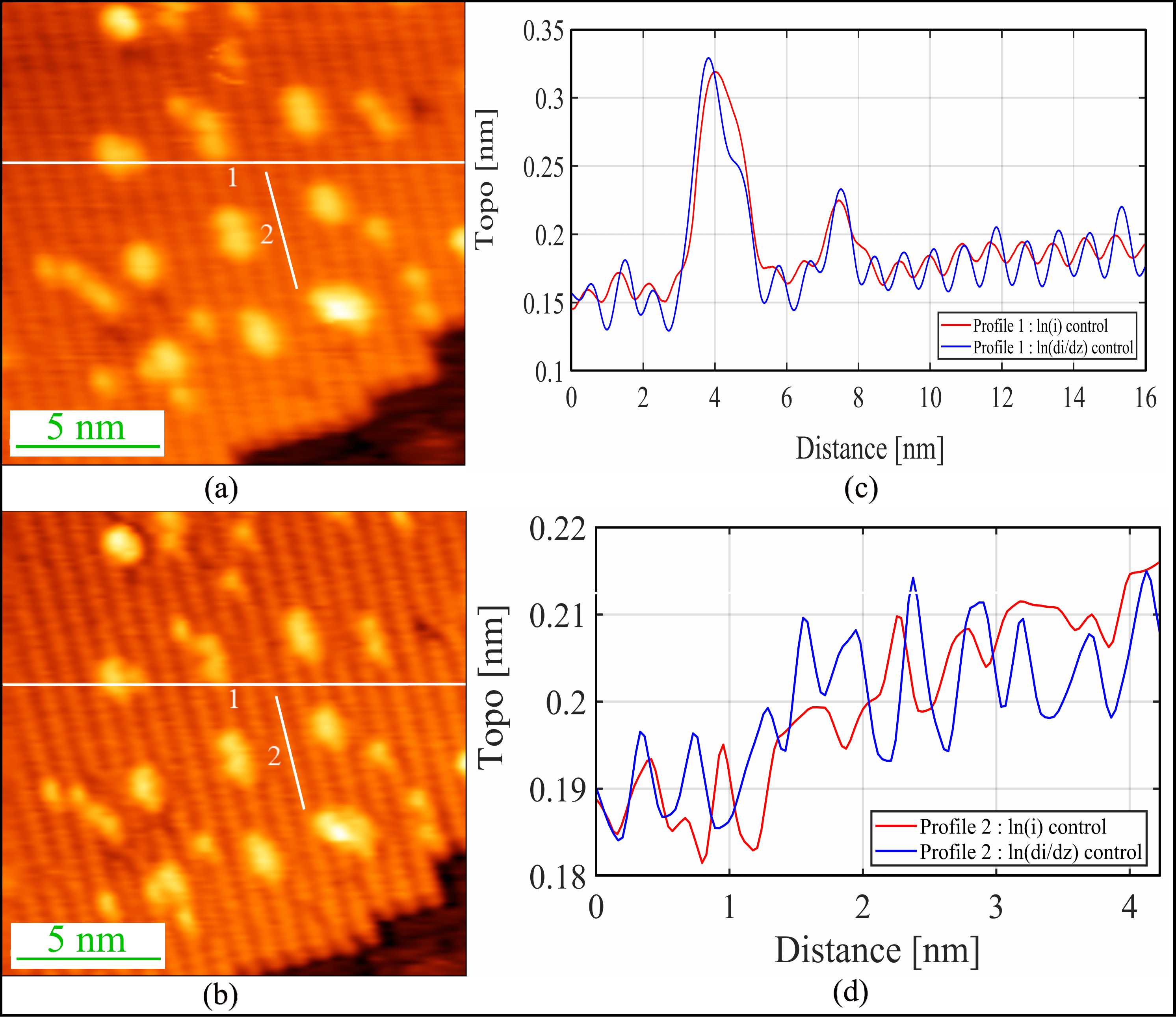}
\caption{The STM images obtained with the constant $ln(Ri)$ and $ln(Rdi/dz)$ feedback loops. (a) Topography image with lithography of an array of $3 \times 3$ dots performed by $ln(Ri)$ feedback loop.  (b) Topography image of same sample area by $ln(Rdi/dz)$ feedback loop. (c) Topography profiles on line 1. (d) Topography profiles on line 2. The imaging parameters are as follows: set-point for constant $ln(Ri)$ = 0.5\,$nA$, set-point for constant $ln(Rdi/dz)$ = 0.5\,$nA/nm$. Lines 1 \& 2 are drawn across the dimer rows and along 11 dimers on both images. For both images: bias voltage= -2.5\,$V$, scan size = 16\,$nm$ $\times$ 16\,$nm$, scan speed = 100\,$nm/sec$, image resolution = 512 $\times$ 512\,$pixels$.  These images are obtained with a ScientaOmicron VT SPM system.}
\label{f:Figure9}
\end{figure}

We also performed the imaging experiment after zooming in on the sample surface to be certain of the improved performance we observed in the first few images with this new feedback control loop. Fig.\,\ref{f:Figure9} a and Fig.\,\ref{f:Figure9} b  were obtained for an image size of $16\;nm\times16\;nm$. We performed lithography to create an array of $3\times3$ dots. In Fig.\,\ref{f:Figure9} a lithography was performed with constant current control, and then an image was also obtained with $ln({Rdi/dz})$ for the same sample area. To better understand the images obtained from both the controllers, we draw two profile lines: 1 across the dimer rows and 2 along the dimers. From  Fig.\,\ref{f:Figure9} c, on profile 1, we can observe distinct dimer rows with better contrast when the loop is closed on $ln(Rdi/dz)$. Also, from  Fig.\,\ref{f:Figure9} d, on profile 2, we can not observe $11$ individual dimers when the loop is closed on $ln(Ri)$. However, we can observe $11$ peaks when the loop is closed on $ln(Rdi/dz)$, which is the expected number of dimers for a length of 4.224\,$nm$.

\subsection{Constant ${di/dz}$ Lithography}
\label{s:Litho_results}

In this section, we demonstrate constant $di/dz$ feedback loop-based HDL. Hydrogen depassivation can occur through two primary mechanisms when the STM works in the constant current mode.
 Above 7\,$V$ bias, depassivation happens through the direct excitation of the Si-H bond, known as field emission (FE) mode. This mechanism yields line widths of approximately 4–5\,$nm$, with rough edges due to partial depassivation.
 Below 4.5\,$V$, HDL operates in what is known as the atomically precise (AP) mode. Here, high current allows for line widths of 0.768\,$nm$, featuring atomically sharp edges. The yield of AP mode lithography is significantly lower than  FE mode lithography ~\cite{lyding1994nanoscale,walsh2009atomic}. However, patterns generated in AP mode demonstrate unparalleled atomic resolution and precision. Linewidth variations are influenced by bias voltage, line dose, and STM tip geometry. Therefore, it is crucial to assess the capability of the new feedback loop for performing AP mode lithography.

To achieve precise control over hydrogen atom desorption along the sample's dimer rows, initial XY-plane piezotube drift is identified and corrected through iterative scanning using Scanz\textsuperscript{\texttrademark} software. After the drift correction, an STM scan captures the surface image and maps the sample's lattice. Subsequently, the tip follows a predefined lithography pattern aligned to the lattice, adjusting tunneling parameters for STM lithography while maintaining closed-loop z-axis control. Tunneling electrons provide energy to desorb hydrogen atoms along this trajectory, leaving a line of silicon atoms with dangling bonds.

\begin{figure}[h!]
\includegraphics[scale=0.38]{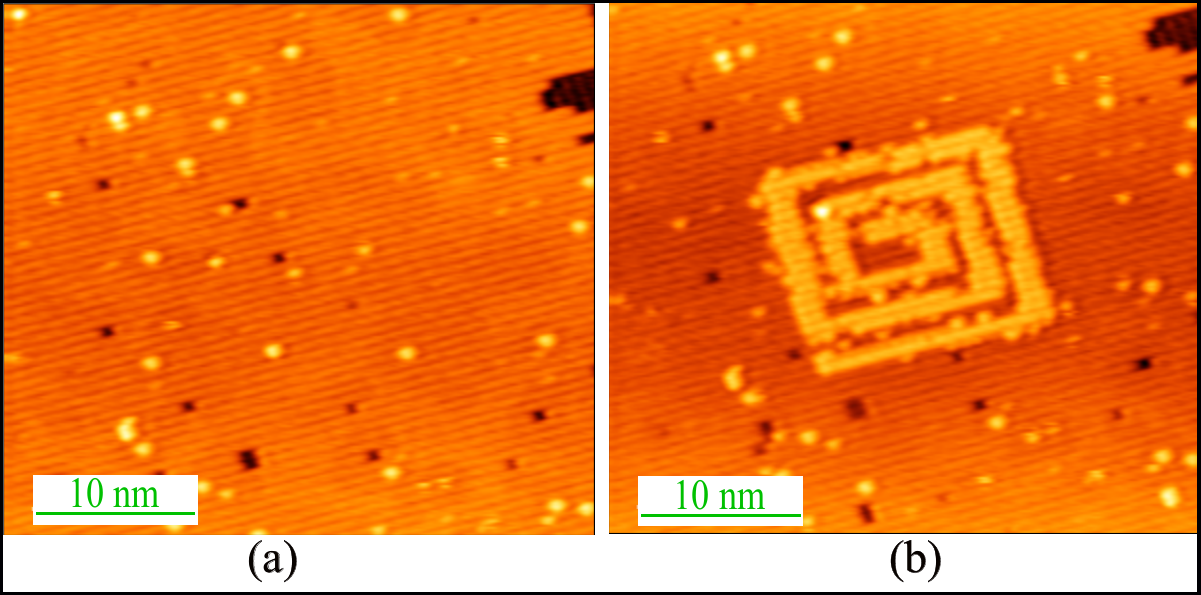}
\caption{STM is operating in constant $di/dz$ mode. (a) Topography images before and  (b) after lithography of the same sample area were captured sequentially. Using a bias voltage of 4\,$V$ and a set-point of 4.0\,$nA/nm$, and tip speed of 10\,$nm/sec$, we created a spiral pattern with three loops on an H-passivated Si\,(100)-2$\times$1 surface. The imaging parameters for both images are as follows: set-point for constant $ln(Rdi/dz)$ = 0.75\,$nA/nm$, bias voltage = -2.5\,$V$, scan size = 32\,$nm$ $\times$ 32\,$nm$, scan speed = 100\,$nm/sec$ and, image resolution = 512$ \times$ 512\,$pixels$. These images are obtained with a ScientaOmicron VT SPM system.}
\label{f:litho1}
\end{figure}

Based on the lithography procedure discussed above, Fig.\,\ref{f:litho1} a is obtained while the loop is closed on $ln(Rdi/dz)$. We performed three-loop spiral lithography with a bias of 4.0\,$V$, set-point of 4.0\,$nA/nm$, and writing speed of 10\,$nm/sec$. The lithography parameters align with those typically employed in conventional STMs to achieve single-dimer-row resolution patterns in AP mode for comparative purposes. The $ln(Rdi/dz)$ feedback loop can produce a clean lithography pattern with atomic resolution, as shown in Fig.\,\ref{f:litho1} b. The lithography spiral pattern is two dimer rows wide along the dimer rows and three dimer rows perpendicular to the dimer rows. We can achieve single dimer row resolution by lowering the bias voltage and identifying the best parameters for a given tip in AP mode.

\begin{figure}[h!]
\includegraphics[scale=0.08]{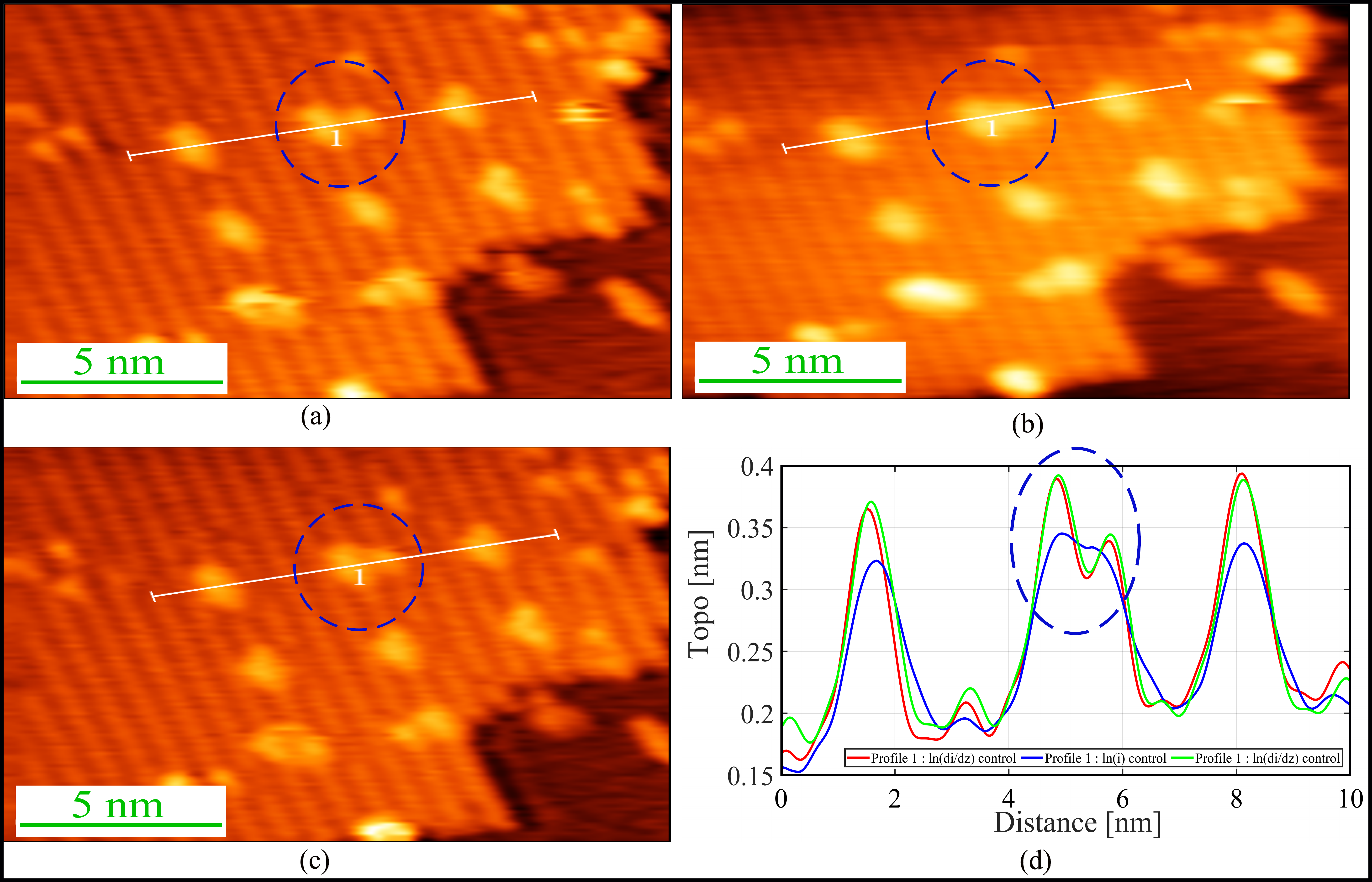}
\caption{The STM is operating in constant $di/dz$ mode. Using a bias voltage of 3.5\,$V$, set-point of 3.0\,$nA/nm$, and tip speed of 7.5\,$nm/sec$, we created an array of $3\times3$ dots on a H-passivated Si\,(100)-2$\times$1 surface. (a) Topography image with lithography performed by $ln(Rdi/dz)$ feedback loop. (b) Topography image of same surface area with $ln(Ri)$ feedback loop. (c) Topography image of same surface area with $ln(Rdi/dz)$ feedback loop. (d) Topography profiles on line 1 for comparison. The imaging parameters for all images are as follows: set-point for constant $ln(Ri)$ = 0.5\,$nA$, set-point for constant $ln(Rdi/dz)$ = 0.5\,$nA/nm$, bias voltage = -2.5\,$V$, scan size = 16\,$nm$ $\times$ 16\,$nm$, scan speed = 100\,$nm/sec$ and, image resolution = 512 $\times$ 512\,$pixels$. These images are obtained with a ScientaOmicron VT SPM system.}
\label{f:litho2}
\end{figure}

\begin{figure*}[t!]
\centering{}
\includegraphics[scale=0.242]{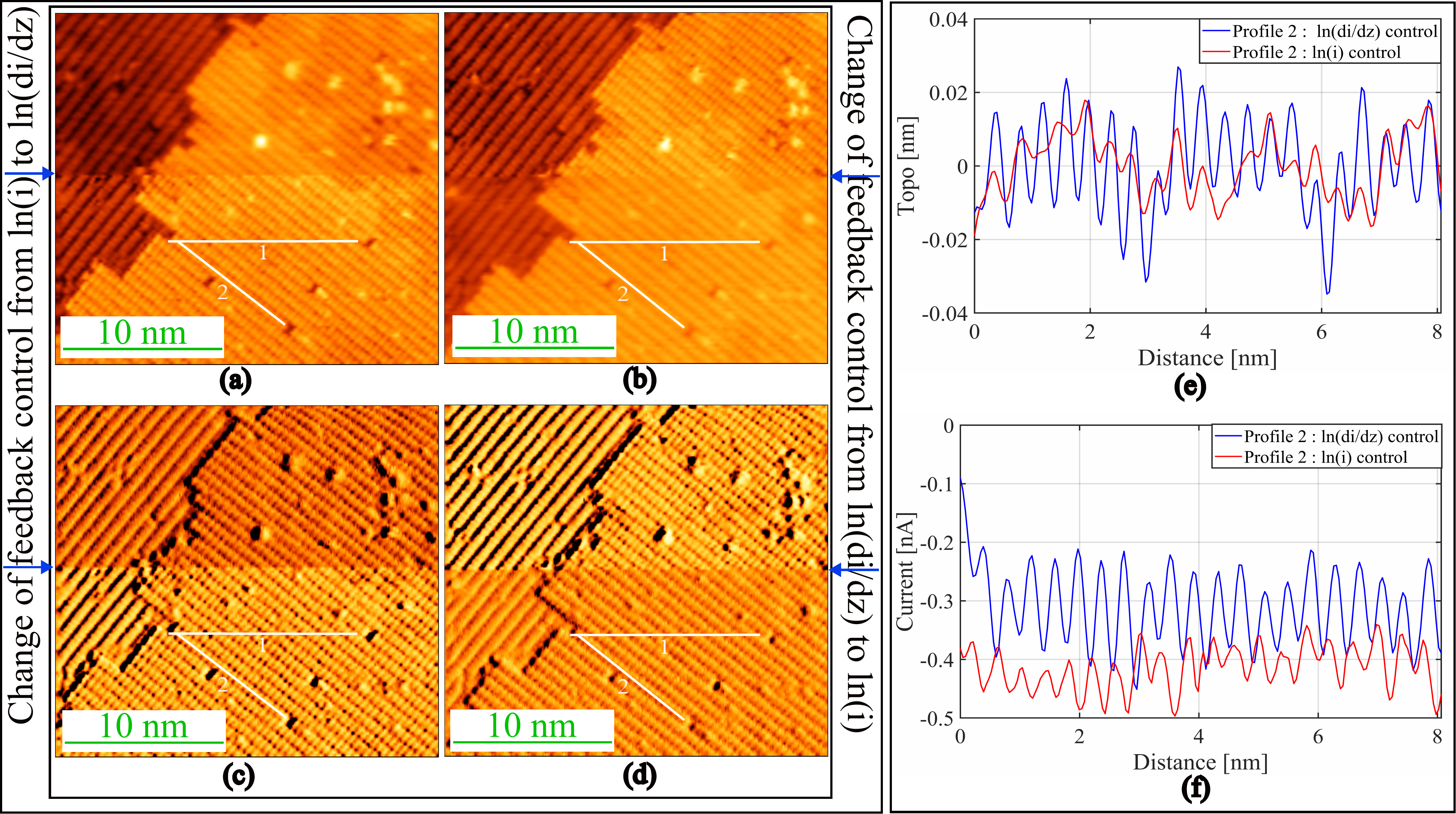}
\caption{ (a) Topography image while feedback loop is transitioning from $ln(Ri)$ to $ln(Rdi/dz)$ measurement. (b) Topography image while feedback loop is transitioning from $ln(Rdi/dz)$ to $ln(Ri)$ measurement. (c) Current image while feedback loop is transitioning from $ln(Ri)$ to $ln(Rdi/dz)$ feedback loop. (d) The current image while the feedback loop is transitioning from $ln(Rdi/dz)$ to the feedback loop of $ln(Ri)$. (e) Topography profiles on line 2. (f) Current profiles on line 2. The imaging parameters for all images are as follows: set-point for constant $ln(Ri)$ = 0.45\,$nA$, set-point for constant $ln(Rdi/dz)$ = 0.26\,$nA/nm$, bias voltage = -2.5\,$V$, scan size = 24\,$nm$ $\times$ 24\,$nm$, scan speed = 100\,$nm/sec$ and, image resolution = 512 $\times$ 512\,$pixels$. These images are obtained with a home-built Lyding scanner-based STM.}
\label{f:compare1}
\end{figure*}

The next lithography experiment is performed to understand the precision of the depassivation process on a $\mathrm{Si(100)-2\times 1:H}$ passivated surface. We performed lithography to create an array of $3\times3$ dots using $ln(Rdi/dz)$ feedback control. We observe from Fig.\,\ref{f:litho2} a that some of the dots in the $3\times3$ dangling bond structure are two dimer rows wide. Also, we can directly compare with Fig.\,\ref{f:Figure9} a, where we created an array of $3\times3$ dots with a feedback loop closed on $ln(Ri)$. Fig.\,\ref{f:litho2} a confirms that lithography performed on $ln(Rdi/dz)$ feedback is on par with lithography performed on $ln(Ri)$ feedback loop.  The set of images in Fig.\,\ref{f:litho2} a-c provided an opportunity to compare the nature of the profile on one of the dangling bonds (in blue circle). For profile 1 corresponding to Fig.\,\ref{f:litho2} a in Fig.\,\ref{f:litho2} d, we observe that there are two peaks, clearly suggesting the formation of two dangling bonds along the adjacent dimer rows. The intensity of the peak suggests the formation of a double-dangling bond or a single-dangling bond during lithography. Fig.\,\ref{f:litho2} b is obtained for the same sample area with the feedback loop closed on $ln(Ri)$. For profile 1 corresponding to Fig.\,\ref{f:litho2} b in Fig.\,\ref{f:litho2} d, we observe a single peak appearing on Fig.\,\ref{f:litho2} b, it appears as one big bright contrast. To establish the sensitivity of this new feedback loop on surface variation, we again closed the loop on $ln(Rdi/dz)$ and obtained Fig.\,\ref{f:litho2} c. We observe two peaks in the corresponding profile in Fig.\,\ref{f:litho2} d and observe two dangling bonds (one big bright contrast and another small bright contrast) in Fig.\,\ref{f:litho2} c.

This lithography experiment and further inspection of the obtained images highlight the capability of the proposed feedback loop to pinpoint the surface features accurately, which is better than conventional feedback control on $ln(Ri)$ given the same tip and imaging conditions.

\subsection{Comparing $ln(Rdi/dz)$ and $ln(Ri)$ feedback mode phenomenon}
\label{s:Litho_results}

We perform another experiment to understand the fundamental phenomenon and the effect of the first term of Eq.\,\ref{11} on imaging conditions. This experiment is performed to rule out any possibility of tip change that may affect/change the contrast of the topography image while scanning. With this experiment, we create a controlled environment to discuss the obtained results and imaging superiority with $ln(Rdi/dz)$ feedback loop. We use the method discussed in Section\,\ref{s:Exp_setup} B and obtain topography and filtered tunneling current images while we change the feedback control loop from $ln(Ri)$ to $ln(Rdi/dz)$ as seen in Fig.\,\ref{f:compare1} a and c. We continue to image and obtain the second set of topography and current images while we now change the feedback loop from $ln(Rdi/dz)$ to $ln(Ri)$ as seen in Fig.\,\ref{f:compare1} b and d. The $ln(Ri)$ feedback loop set-point is 0.5\,$nA$. When we switch the feedback loop to $ln(Rdi/dz)$ mode, the set-point is the $di/dz$ value obtained when the feedback loop is closed on $ln(Ri)$. We observe a transition on topography and current images as we switch the feedback loop, as marked by the blue arrow in Fig.\,\ref{f:compare1} a-d. No tip change is observed besides the change in imaging condition due to a change in the feedback loop.

The superior contrast obtained in topography images from $ln(Rdi/dz)$ feedback loop while we switch from $ln(Ri)$ to $ln(Rdi/dz)$ (Fig.\,\ref{f:compare1} a (bottom)) or {\it vice versa} (Fig.\,\ref{f:compare1} b (top)) is visible in both sets of images given the same tip conditions. As we discussed in Section\,\ref{sec:modulation}. C, this higher resolution is because the tip undergoes smaller height variations during a scan due to electronic disturbances but a similar movement due to surface height variations. This increases the sensitivity of the tip to surface variations when the STM operates in constant $di/dz$ mode. Two profile lines are drawn on the images to understand the phenomenon. First, across the dimer rows (Profile 1) and second, along the dimer rows (Profile 2). We consider the profile $2$ on topography and current images to investigate and obtain differences in both feedback loops. We observe that besides the difference in visual contrast between  Fig.\,\ref{f:compare1} a and b (bottom), we can count 21 dimers on profile 2 in Fig.\,\ref{f:compare1} a as compared to profile 2 in Fig.\,\ref{f:compare1} b. This behavior is also highlighted in Fig.\,\ref{f:compare1} e, where the topography profile is scaled to overlap and highlight the height difference obtained from $ln(Ri)$ and $ln(Rdi/dz)$. Comparing profile 2 on current images from Fig.\,\ref{f:compare1} c and d (bottom), we observe in Fig.\,\ref{f:compare1} f, the average tunneling current decreases from -0.41\,$nA$ to -0.31\,$nA$ when we change the feedback loop closed on $ln(Ri)$ to $ln(Rdi/dz)$ respectively. The decrease in average tunneling current implies an increase in tip-sample distance, emphasizing the impact of the first term of Eq.\,\ref{11}. We also observe similar tunneling current behavior on profile 1. Overall, we can conclude that the controller has to make less effort in imaging while the loop is closed on $ln(Rdi/dz)$. This feedback control system can image with a superior contrast from increased tip-sample distance due to a reduction in output disturbance, $\ln((- 1.025\sqrt{\varphi}\;f(\sigma, V_b)))$ for H-passivated Si\,(100)-2$\times$1 surfaces.

 \section{Conclusion}
\label{s:Conclusion4}

In this work, we presented a novel constant $di/dz$ mode scanning tunneling microscopy (STM) technique. We demonstrated improved image contrast by modulating the controller output to regulate the tip-sample distance compared to conventional constant current mode STM. 
We successfully applied this new control mode to a variety of STM experiments, demonstrating its stability and reliability. The proposed feedback loop offers several advantages, particularly for future STM imaging and high-throughput hydrogen depassivation lithography applications involving arrays of MEMS STM tips. The ability to independently control each tip and multiplex the digital $di/dz$ signals allows for efficient parallel operation without increasing hardware costs. Overall, the constant $di/dz$ mode STM presents a promising alternative to conventional STM techniques, offering improved performance and scalability.

\section{\label{sec:acknow}Acknowledgments}
This work is partially supported by the US Department of Energy under grant number DE-SC0020827 and partially by the UTD Quantum Center. The authors wish to thank Zyvex Labs members for assisting in setting up the experimental test beds.

\section{\label{sec:AuthDec}Author declarations}

\subsection{\label{sec:AuthDec}Conflict of interest}

The authors have no conflicts to disclose.

\section{\label{sec:data}Data availability}

The data that supports the findings of this study are available within the article.

\appendix

\section{Appendix} 
\label{app:subsec}

\begin{figure}[h!]
\centering{}
\includegraphics[scale=0.2]{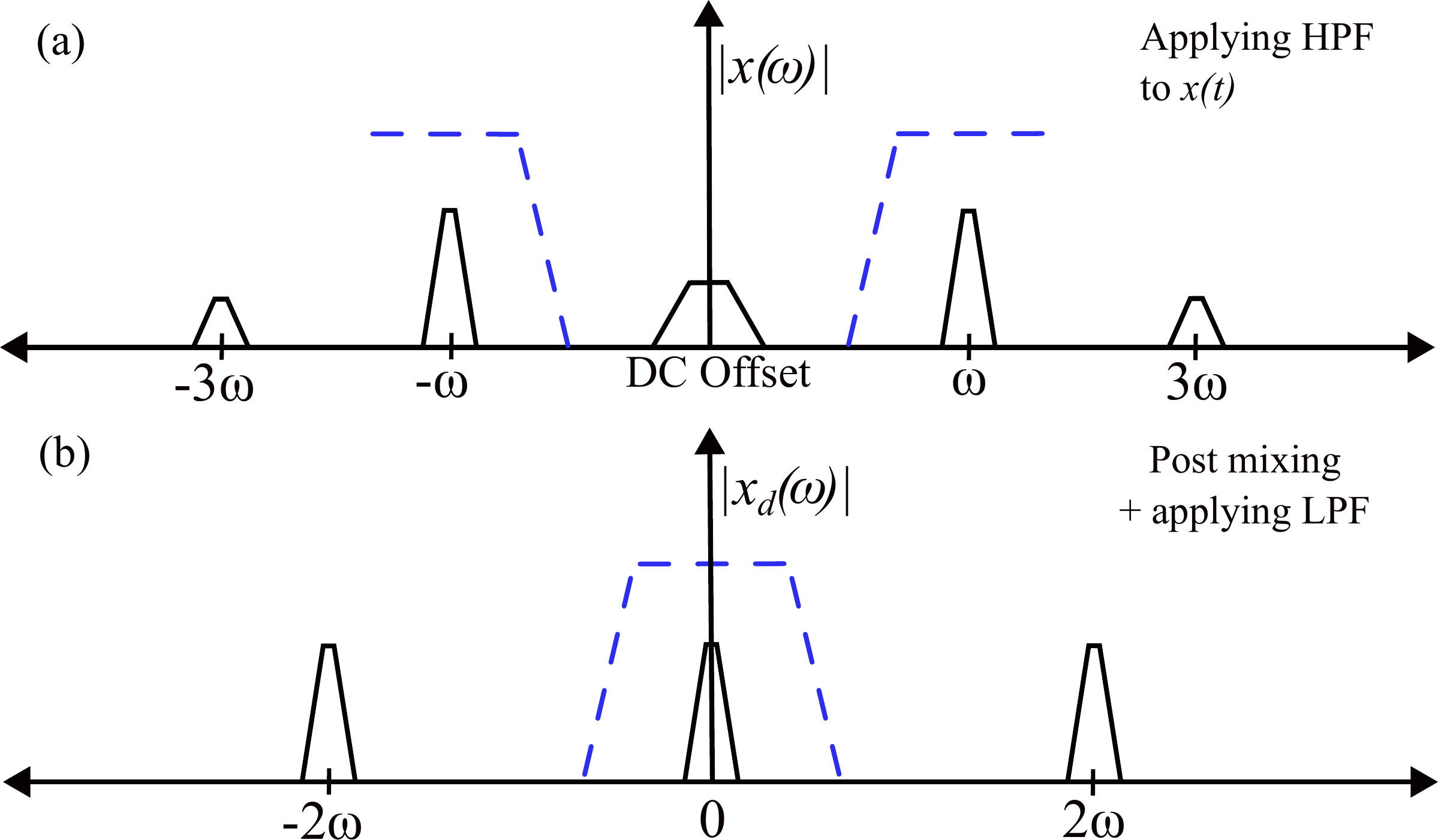}
\caption{(a) Illustrative double-sided amplitude spectrum of the signal after high pass filter (HPF) and (b) after mixing and low pass filtering.}
\label{f:LIAFS}
\end{figure}

Consider a scenario in which the measured signal comprises a spectrum of various frequency components, represented by the Fourier series as:
\begin{equation} \label{eq3}
x(t) = a_{0}(t) +  \sum_{n=1}^{\infty} a_{n} sin(\omega_{n}t + \phi_{n}) + n      
\end{equation}
where $n$ is a measurement noise. We need to estimate the signal amplitude, $a_r$ at the frequency of interest $\omega_n = \omega_r$ in the measured signal, $x(t)$. The LIA algorithm starts by AC coupling the signal $x(t)$ using a high-pass filter to eliminate any DC bias, flicker noise, and low-frequency components of tunneling current, collectively represented by $a_0(t)$ in Eq.\,\ref{eq3} and shown in Fig.\,\ref{f:LIAFS} a, for enhanced measurement accuracy. We then modulate the high pass filtered signal, $x_f(t)$ with  $x_r(t) = sin(\omega_{r}t)$ as show in  Fig.\,\ref{f:LIA}. The signal obtained after mixing these signals is:

\begin{equation} \label{eq4}
\begin{split}
x_d(t) & = x_f(t)\cdot x_r(t)\\
 & = \sum_{n=1}^{\infty} a_{n} sin(\omega_{n}t + \phi_{n})\; sin(\omega_{r}t) + n\cdot  sin(\omega_{r}t) \\
 & =  \underbrace{n\cdot sin(\omega_{r}t)}_{\RN{1}} + \underbrace{\frac{a_{r}}{2}[cos\phi_{r} - cos(2\omega_rt + \phi_r)]}_{\RN{2}\; for \; \omega_n = \omega_r} +\\
 & \underbrace{ \frac{a_n}{2} \Biggr[\sum_{n=1,n \neq r}^{\infty}(cos(\omega_n - \omega_r)t + \phi_n) -  (cos(\omega_n + \omega_r)t + \phi_n) \Biggr]}_{\RN{3}}
\end{split}
\end{equation}

\begin{equation} \label{eq5}
\begin{split}
x_q(t) & = x_f(t)\cdot x_r(t + \frac{\pi}{2})\\
 & = \sum_{n=1}^{\infty} a_{n} sin(\omega_{n}t + \phi_{n})\; cos(\omega_{r}t) + n\cdot cos(\omega_{r}t) \\
 & = n\cdot cos(\omega_{r}t) + \frac{a_{r}}{2}[sin\phi_{r} + sin(2\omega_{r}t + \phi_r)] +\\
 &\frac{a_n}{2} \Biggr[\sum_{n=1,n \neq r}^{\infty}(sin(\omega_n - \omega_r)t + \phi_n) +  (sin(\omega_n + \omega_r)t + \phi_n) \Biggr]
\end{split}
\end{equation}

Eqs. \,\ref{eq4} and \,\ref{eq5} reveal that in addition to the components of interest (first term of $\RN{2}$), the modulation process generates other mixing products at integer multiples of the modulation frequency (second term of $\RN{2}$). These higher-frequency components (second term of $\RN{2}$, $\RN{3}$ term and noise ($\RN{1}$ term)) are removed by a low-pass filter (LPF) as illustratively shown in Fig.\,\ref{f:LIAFS} b. Using the output $y_{d_{dc}}  = a_r cos\phi_r/2$ and $y_{q_{dc}} = a_r sin\phi_r/2$, we can obtain the amplitude $a_r$ of the signal at $\omega_r$ frequency.
\begin{equation} \label{eq6}
A = 2\cdot\sqrt{{y_{d_{dc}}}^2 + {y_{q_{dc}}}^2} = a_r
\end{equation}
We modulate the physical parameter of interest to apply the lock-in amplifier practically. This parameter influences the system under study to respond at the modulation frequency to produce the signal $x(t)$ as shown in Fig.\,\ref{f:LIA}. The primary measured signal in STM is the tunneling current, $i$, and the other measured signals are controller output and bias voltage, which the user can set. We control the parameter $p$, which could be any other measured signal of the system. This stimulus varies sinusoidally around an average value $\bar{p}$ at the frequency $\omega$  with amplitude $A_m$ such that, $x(t) = f(p(t))$, where $p = \bar{p} + A_m sin(\omega t)$. The Taylor series expansion of $x$ around the parameter $p$ is as follows:
\begin{equation} \label{eq7}
\begin{split}
x(t) & = f(p(t))\\
 & = f(\bar{p}) + \frac{\partial f}{\partial p}\Bigg|_{p = \bar{p}} \frac{(p- \bar{p})}{1!} 
  + \frac{\partial^{2}f}{\partial p^{2}}\Bigg|_{p = \bar{p}} \frac{(p- \bar{p})^2}{2!}\\
& + \frac{\partial^{3}f}{\partial p^{3}}\Bigg|_{p = \bar{p}} \frac{(p- \bar{p})^3}{3!} + \mathcal O(A^2)
\end{split}
\end{equation}
The signal, $x(t)$ comprising of $\omega, 2\omega, 3\omega$ frequency term is passed through the LIA. The LIA picks up the component at the reference frequency, $\omega_r$. If the frequency of interest is $\omega_r= \omega$, then LIA will output the amplitude corresponding to the $\partial f/\partial p$ term. As long as the amplitude of the modulation frequency is small enough, this relation is linear and yields $\partial f/\partial p$. Otherwise, the output becomes distorted, and the outputs are no longer proportional to $\partial f/\partial p$. Similarly, if the frequency of interest is $2\omega$, then the amplitude is proportional to $\partial^2f/\partial p^2$, ignoring the higher order terms.

\bibliography{References}

\balance

\end{document}